\documentclass[]{jfm}

\usepackage{graphicx,caption}
\usepackage{newtxtext}
\usepackage{newtxmath}
\usepackage{natbib}
\usepackage{hyperref}
\usepackage{textcomp}
\usepackage{gensymb}
\hypersetup{
    colorlinks = true,
    urlcolor   = blue,
    citecolor  = black,
}

\newcommand{\RomanNumeralCaps}[1]
\linenumbers

\newcommand*{\xdash}[1][3em]{\rule[0.5ex]{#1}{0.55pt}}
\newcommand\Ra{\mbox{\rm{Ra}}}
\newcommand\Ha{\mbox{\rm{Ha}}}

\newcommand{\zstroke}{%
  \text{\ooalign{\hidewidth -\kern-.3em-\hidewidth\cr$z$\cr}}%
}
\newcommand{\sz}{\ensuremath{\scalebox{1}{z}
\kern-.4em\raisebox{-.0em}{$\xdash[.4em]\hspace{.1em}$}}}

\title{Subcritical transition and multistability in liquid metal magnetoconvection with sidewalls}

\author{Matthew McCormack\aff{1},
    Andrei Teimurazov\aff{2},
    Olga Shishkina\aff{2}
\and Moritz Linkmann\aff{1}
\corresp{\email{moritz.linkmann@ed.ac.uk}}}

\affiliation{\aff{1}School of Mathematics and Maxwell Institute for Mathematical Sciences, University of Edinburgh, UK\aff{2}Max Planck Institute for Dynamics and Self-Organization, 37077 Göttingen, Germany}

\begin{document}
\maketitle

\begin{abstract}
The motionless conducting state of liquid metal convection with an applied
vertical magnetic field confined in a vessel with insulating side walls becomes linearly unstable to wall modes through a supercritical pitchfork
bifurcation. Nevertheless, we show that the transition
proceeds subcritically, with stable finite-amplitude
solutions with different symmetries existing at parameter values beneath this linear stability threshold.
Under increased thermal driving, the branch born from the linear instability becomes unstable and solutions are attracted to the most subcritical branch, which follows a quasiperiodic route to chaos.
Thus, we show that the transition to turbulence is controlled by this subcritical branch and hence, turbulent solutions have no connection to the initial linear instability.  This is further quantified by observing that the subcritical equilibrium solution sets the spatial symmetry of the turbulent mean flow and thus, organises large-scale structures in the turbulent regime. 
\end{abstract}

\begin{keywords}
Magnetoconvection, Instability, Nonlinear Dynamical Systems 
\end{keywords}


\vspace{-30pt}
\section{Introduction}
Liquid metal magnetoconvection (MC), where an electrically conducting fluid is subjected to a time-independent magnetic field and an imposed temperature gradient, is relevant to engineering applications such as liquid-metal cooling systems for nuclear fusion reactors \citep{Davidson1999}, geo-astrophysical systems such as flows in liquid-metal planetary cores \citep{Jones2011}, and associated laboratory experiments \citep{stefani2024liquid}.  Despite its relevance, however, a number of uncertainties persist regarding the transition to turbulence in this system, especially in confined containers relevant to experiments and engineering applications.  Here, we focus on the case where the uniform applied magnetic field is orientated in the vertical direction, directly opposing gravity.  In contrast to magnetoconvection in a laterally periodic/infinite plane layer, where a linear instability of the motionless conducting state classically gives rise to domain-filling arrays of convection rolls \citep{chandrasekhar1961hydrodynamic,Busse1982}, wall-localised onset modes, named \emph{wall modes}, are observed when sidewalls are introduced to the system with a sufficiently strong magnetic field.  Wall modes observed in direct numerical simulations (DNS) \citep{Liu2018,akhmedagaev2020turbulent,xu2023transition,McCormack_Teimurazov_Shishkina_Linkmann_2023,wu2025flow} and experiments \citep{zurner2020flow,xu2023transition} have been thought to originate from a linear instability of the conducting state, originating from the analytic-numerical hybrid approach of \cite{houchens2002rayleigh}, and the asymptotic (large magnetic field limit) linear stability analysis of \cite{busse2008asymptotic}.  However, recent work suggests discrepancies between linear stability theory, simulations and experiments in the measurement of the onset of convection \citep{xu2023transition}, and short-time simulations displaying exponential growth from a perturbed conducting state were observed to have a different modal structure than converged wall mode equilibria at similar parameter values \citep{McCormack_Teimurazov_Shishkina_Linkmann_2023}.  In this work, we show for the first time that multiple stable wall mode solutions exist at identical operating conditions, with fundamental consequences for the transition to turbulence in this flow.  

In a general fluid dynamics setting, systems typically either follow a supercritical or subcritical transition to turbulence.  A supercritical transition occurs when the laminar flow undergoes successive supercritical bifurcations, transitioning the flow to an increasingly spatio-temporally complex state.  Each transition point is clearly identifiable in parameter space, and may be found using a linear stability analysis due to the local nature of the instabilities.  Such behaviour has been identified in many flow configurations such as Taylor-Couette flow (TCF) \citep{taylor1923viii,taylorCouette-swinney-1975,taylorCouette-swinney-1987,taylorCouetteBook}, and Rayleigh--B\'enard convection (RBC) \citep{rayleigh1916lix,Malkus_Veronis_1958,chandrasekhar1961hydrodynamic,libchaber1982period,PhysRevA.44.8103,bodenschatz2000rbc,PhysRevE.81.036321,oteski2015quasiperiodic}.
In other cases, however, the transition proceeds subcritically, where finite-amplitude disturbances trigger the transition despite a linearly stable laminar flow.  In this case, non-trivial nonlinear solutions must be identified beneath the linear stability threshold to understand the transition.  This leads to a different phenomenology, featuring hysteresis and large, unpredictable changes in the system response triggered at a range of parameter values depending on the supplied disturbance. 
In fluid dynamics and related fields, subcritical transitions and hysteretic effects occur in a wide range of systems. They are prominent concerning the transition to turbulence in wall-bounded shear flows of neutral
\citep{Kerswell2005,Eckhardt2007, 
eckhardt2018transition, 
Avila2023, Hof2023} and electrically conducting fluids (for comprehensive review see \citet{Zikanov2014}), in the transition to elasto-inertial and 
elastic turbulence for flows with straight streamlines \citep{Morozov2007, Bonn2011, Pan2013, 
lellep2024elastic}, and for types of (approximately) homogeneous turbulence \citep{Linkmann2015}. Finite-amplitude transitions between different turbulent  states, such as those characterised by the presence or absence of a condensate, occur in continuum models describing certain regimes of active turbulence 
\citep{Linkmann2019, Linkmann2020a},
two-dimensional turbulence depending on the type of forcing \citep{Linkmann2020b, Gallet2024} and in quasi-two-dimensional turbulence occurring in rotating systems or thin fluid layers \citep{favier2019subcritical, deWit2022a, deWit2022b, yokoyama2017}. 
Moreover, there are more complex scenarios, such as the transition to the ultimate regime of thermal convection \citep{Roche2020, lohse2023ultimate, lohse2024ultimate,shishkina2024ultimate}
which occurs through a subcritical transition from laminar to turbulent boundary layers, in presence of turbulent flow in the bulk.

Liquid metal MC (\emph{i.e.} large magnetic diffusivity) transitions supercritically in a laterally periodic layer \citep{chandrasekhar1961hydrodynamic,Busse1982,proctor1982magnetoconvection,weiss2014magnetoconvection}, and the motionless conducting state undergoes a supercritical bifurcation when laterally confined between sidewalls \citep{houchens2002rayleigh,busse2008asymptotic,bhattacharya2024wall}.  Despite this, we show that non-trivial nonlinear equilibrium solutions exist beneath the wall mode linear stability threshold, and that the transition proceeds on this subcritical branch of solutions.  We further observe that this has a fundamental impact on the organisation of large-scale structures in the turbulent regime.  This implies that (a), the transition to turbulence and turbulent solutions in closed vessel experiments are fundamentally different to the laterally unbounded geo-astrophysical flows they often intend to replicate in nature, and (b), we must change the way we think about the transition and the formation of large-scale structures in convective and potentially other non-linear pattern-forming systems subject to external forces, even when linear stability theory predicts a supercritical transition.
 
\section{Formulation}
An incompressible flow of viscous and electrically conducting fluid is driven by an imposed vertical temperature difference, $\delta T$,
and a constant vertical magnetic field $\boldsymbol{B} = B_0\boldsymbol{e}_z$.  The equations of motion under the quasistatic approximation (magnetic Reynolds $\rm{Rm} = \it{U} \ell/\eta \ll \rm1$ and magnetic Prandtl number $\rm{Pm} = \nu/\eta \ll 1$, where $U$ and $\ell$ are characteristic velocity and length scales, $\nu$ is the kinematic viscosity and $\eta$ is the magnetic diffusivity) and the Oberbeck-Boussinessq approximation are
\vspace{-5pt}
\begin{subequations}
\label{eq:governing}
\begin{alignat}{1}
    \partial_t\boldsymbol{u} + \boldsymbol{u}\bcdot\bnabla\boldsymbol{u}+\bnabla p &= \sqrt{\frac{\Pr}{\Ra}} \big[\Delta\boldsymbol{u} + \Ha ^2(\boldsymbol{j}\times\boldsymbol{e}_z)\big] + T\boldsymbol{e}_z, \\
    \partial_t T + \boldsymbol{u}\bcdot\bnabla T &= \frac{1}{\sqrt{\Ra\Pr}}\Delta T, \\
    \bnabla \bcdot \boldsymbol{u} = 0, \quad 
    \boldsymbol{j} = -\bnabla\phi &+ (\boldsymbol{u}\times\boldsymbol{e}_z), \quad
    \Delta\phi = \bnabla\bcdot(\boldsymbol{u}\times\boldsymbol{e}_z),
\end{alignat}    
\end{subequations}
where $\boldsymbol{u} = [u,v,w]^T$ is the velocity, $T$ the temperature, $p$ the kinematic pressure, $\boldsymbol{j}$ the electric current density, and $\phi$ the electric field potential.     Quantities have been non-dimensionalised using the height of the fluid layer $H$,
the temperature difference $\delta T\equiv T_+-T_- > 0$, where $T_+$ and $T_-$ are the temperatures at the bottom and top plates respectively, the free-fall velocity $u_{f\!f}\equiv(\alpha gH\delta T)^{1/2}$, 
the free-fall time $t_{f\!f}\equiv H/u_{f\!f}$, 
the pressure $\rho u_{f\!f}^2$ and
the external magnetic field strength $B_0$, where $\alpha$ is the thermal expansion coefficient and $g$ the acceleration due to gravity.
The control parameters are the Rayleigh number $\Ra$, Prandtl number $\Pran$, and Hartmann number $\Ha$,
\vspace{-5pt}
\begin{eqnarray}
\Ra\equiv~\dfrac{\alpha g \,\delta T\, H^3}{\kappa \nu}, \qquad \Pran\equiv~\dfrac{\nu}{\kappa}, \qquad \Ha\equiv~B_0 H \sqrt{\dfrac{\sigma}{\rho \nu}} \ , \label{gov_par}   
\end{eqnarray}
where 
$\kappa$ is the thermal diffusivity,
$\sigma$ the electrical conductivity,
 and $\rho$ the mass density.  We apply no-slip boundary conditions (BCs)
 for the velocity at all boundaries $\boldsymbol{u}=0$, and adiabatic BC at the side walls, $\partial T/\partial \boldsymbol{n}=0$, where $\boldsymbol{n}$ is the vector orthogonal to the surface.  
All solid boundaries are considered electrically insulating $\partial \phi/\partial \boldsymbol{n}=0$.
The spatial domain $\Omega\subset\mathbb{R}^3$ is a cube with width $W$ and length $L$, i.e. $H=W=L$, and is resolved on non-uniform grids using the MC extension of {\sc goldfish} \citep{kooij2018comparison,reiter2021crossover,reiter2022flow,McCormack_Teimurazov_Shishkina_Linkmann_2023,Teimurazov_2024}, which uses a third-order Runge-Kutta time integration scheme and a fourth-order finite-volume discretisation.  The coordinates of the grid points, discretizing the interval $[0,1]$, are defined as $x_k = [1-(2\tilde{x}_k/(\tilde{x}_1 - \tilde{x}_n))]/2$ for $k = 1, \ldots, n$, where $\tilde{x}_k$ are generalized Chebyshev nodes $\tilde{x}_k = \cos\big((2k + 2k_{clip}-1)\pi/(2n + 4k_{clip})\big)$ with uniformity parameter $k_{clip} \in \mathbb{N}$.
In our DNS, we use at least $220^2$ points in the cross-plane direction with $k_{\rm clip} = 10$, and 350 points in the vertical direction with $k_{\rm clip} = 1$ in the vertical direction.  With this grid, spatial flow fluctuations are resolved to less than 2 Kolmogorov microscales for all simulations used in the stability analysis, with no less than 9 points in the Hartmann boundary layer which forms on the top/bottom boundaries of the domain.  Here, the Hartmann layer thickness is defined as $\delta_{\nu}\approx1/\Ha$ which uses the prefactor measured by \citet{Teimurazov_2024}.  This layer first forms near onset in the boundary layer underneath the wall modes and thus, resolving this layer is critical to correctly analysing the stability of the solutions. Simulations with $\Ra\geq10^8$ have been resolved on a finer $250^2\times400$ grid to explore the long-time behaviour of the system, the results of which have been validated with shorter runs using a $350^2\times600$ grid. We set $\Pran=0.025$ for liquid metals such as Gallium-Indium-Tin, and fix the Hartmann number at $\Ha=500$ or $\Ha=1000$, leaving the Rayleigh number $\Ra$ as our bifurcation parameter.

\vspace{-10pt}
\section{Multistability of nonlinear wall mode equilibria}
\subsection{Linear stability and the linear onset branch}
\begin{figure}
 \captionsetup{width=\columnwidth}
    \includegraphics[width = 0.95\columnwidth]{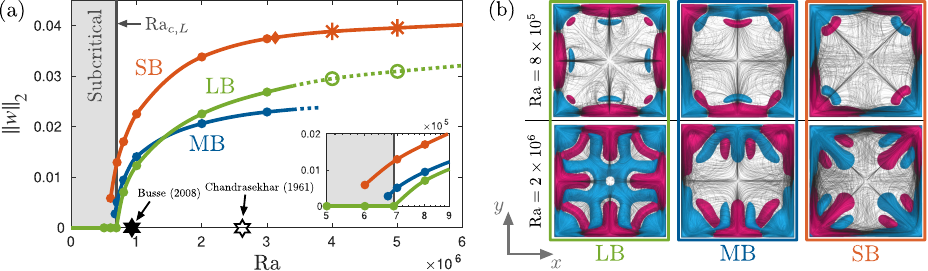}
    \caption{(a) Bifurcation diagram at $\Ha=500$ showing stable/unstable equilibria (filled/open markers), limit cycles (diamonds), invariant tori (asterisks) on the linear onset branch (LB) stemming from the linear instability at $\Ra_{c,L}$, the mixed symmetry branch (MB) and the subcritical branch (SB).  The mean of time-dependent solutions is shown.  The grey area represents $\Ra<\Ra_{c,L}$.   (b) Upflow (pink) and downflow (blue) vertical velocity isosurfaces and streamlines (black) of solutions on the various branches shown from the top view at $\Ha=500$ at $\Ra=8\times10^5$ (top) and $\Ra=2\times10^6$ (bottom) with $w=\pm0.005$ and $\pm0.01$, respectively.}
    \label{fig:bifurc-diag}
 \end{figure}

Although a strong magnetic field can completely suppress convection, the motionless conducting state becomes linearly unstable with sufficient thermal driving 
in an infinitely wide and long domain, with the critical Rayleigh number $\Ra_c \sim \Ha^2$ \citep{chandrasekhar1961hydrodynamic} (shown by the open black star in Fig.~\ref{fig:bifurc-diag}(a)).  
However, including insulating sidewalls results in 
a different linear instability giving rise to wall modes beneath the onset of bulk convection
\citep{houchens2002rayleigh,busse2008asymptotic,bhattacharya2024wall}, even in large-aspect ratio domains (also with finite sidewall conductivity).  The critical Rayleigh number for a sidewall in a semi-infinite domain with free-slip top and bottom boundaries is $\Ra_c \sim \Ha^{3/2}$ for large $\Ha$ \citep{busse2008asymptotic} (shown by the filled black star in Fig.~\ref{fig:bifurc-diag}(a)).  
Since no stability theory has been performed for the domain and boundary conditions we consider, or any other fully confined domain, we first perform a linear stability analysis for the conducting state.  The linear stability threshold is obtained by perturbing the conducting state with random noise of magnitude $O(10^{-10})$ and scanning $\Ra$ to find an approximate threshold.  Two equilibria are converged on each side of the threshold with $|\partial_t E_k| = O(10^{-14})$ where $E_k = \frac{1}{2}\int_\Omega(\boldsymbol{u}\bcdot\boldsymbol{u}) \,\mathrm{d} \boldsymbol{x}$, ensuring agreement between the fastest growing/slowest decaying modes.  This modal structure (similar to Fig.~\ref{fig:bifurc-diag}(b) (LB) at
$\Ra=8\times10^5$) is used as the spatial eigenmode to bound the linear stability threshold $\Ra_{c,L}$ by bisection. 
For $\Ha=500$ we have $6.89 \times 10^5 <
\Ra_{c,L} < 6.90 \times 10^5$ and at $\Ha=1000$ we have $ 1.83 \times 10^6 <
\Ra_{c,L} < 1.84 \times 10^6$.  Locally, based on these two points,
$\Ra_{c,L}\sim\Ha^{1.412\pm0.005}$, where the error corresponds to the uncertainty in $\Ra_{c,L}$ at each $\Ha$.  Although this is a local measurement of the scaling at finite $\Ha$, it is close to the asymptotic result for a single sidewall obtained by \cite{busse2008asymptotic} where $\Ra_{c}\sim\Ha^{3/2}$ at leading order.  The stable equilibria born from the linear instability (linear onset branch (LB) in the bifurcation diagram in Fig. \ref{fig:bifurc-diag}(a)) exhibit a discrete 4-fold rotational symmetry in the vertical velocity field (Fig. \ref{fig:bifurc-diag}(b) (LB)), and is thus invariant under the rotation $r: rw(x,y,z) = \mathcal{R}_{\pi/2} w(x,y,z),$ where $\mathcal{R}_\theta$ represents a rotation $\theta$ about $\boldsymbol{e}_z$.  The solution additionally has reflection symmetries in the $x$-$y$ plane (about the $x$ and $y$ axes, and about the two diagonals).  Thus, these symmetries may be described by a group isomorphic to the dihedral group $D_4$, generated by $r$, the identity $\mathrm{id}$ and the reflection $s$ about the $x=y$ diagonal, which we will denote $D_4^+ = \{\mathrm{id}, r, r^2, r^3, s, sr, sr^2, sr^3\}$.
These solutions have thin convective wall modes with a two-layer
structure near the sidewalls, with convection almost completely suppressed in
the bulk.  Each sidewall features two counter-rotating rolls, which lead to the
spatially periodic pattern of up/down flows shown in the vertical velocity
isosurfaces in Fig. \ref{fig:bifurc-diag}(b) (LB).  A secondary solution can be obtained by changing the direction of the rolls and flipping the vertical coordinate $ \sz: \sz w(x,y,z)= -w(x,y,-z)$, and thus, the bifurcation is a supercritical pitchfork, which has broken the $\mathbb{Z}_2$ Boussinesq symmetry of the $D_4 \times\mathbb{Z}_2 = D_4^+ \times \{\mathrm{id},\sz\}$ conducting state.  Increasing $\Ra$ results in protrusions growing from the wall modes into the bulk of the flow, similar to other observed solutions \citep{Liu2018,xu2023transition,McCormack_Teimurazov_Shishkina_Linkmann_2023}, and the equilibria eventually lose
stability between $ 3 \times 10^6 < \Ra <4\times10^6$.  Since solutions on this
branch are only weakly unstable up to $\Ra\approx10^7$, we have estimated the continuation of the unstable branch in Fig. \ref{fig:bifurc-diag}(a).

 \vspace{-10pt}
\subsection{Subcritical branch}
Having determined the threshold for a linear instability, we now assess the nonlinear stability of the conducting state and search for subcritical solutions.
We note that the solutions on LB have a different spatial symmetry than other wall mode solutions in the same domain \citep{McCormack_Teimurazov_Shishkina_Linkmann_2023}.
Thus, to try to construct a different branch of solutions, we initialise
from a solution obtained far from the threshold of the linear instability ($\Ra=5\times 10^7$) at the given $\Ha$, which displays chaotic dynamics, and then slowly reduce $\Ra$ until we encounter equilibria.  The obtained solutions 
again feature wall-mode structures but indeed have a different symmetry 
in the vertical velocity field (the same as in \cite{McCormack_Teimurazov_Shishkina_Linkmann_2023}), now invariant under the rotation $\sz r: (\sz r) w(x,y,z) = -\mathcal{R}_{\pi/2} w(x,y,-z)$ which features a reflection in sign and vertical coordinate (Fig. \ref{fig:bifurc-diag}(b) (SB)).  This state still exhibits reflection symmetries about the diagonals in the $x-y$ plane, but also along the $x$ and $y$ axes, this time accompanied by a reflection in sign and vertical coordinate.  Thus, the symmetry group is again isomorphic to $D_4$ but now with the different generators, $r_- = (\sz r), \mathrm{id}$, and $s$, which we will denote $D_4^- = \{\mathrm{id}, r_-, r_-^2, r_-^3, s, sr_-, sr_-^2, sr_-^3\}$.  Importantly, we have found that this
solution branch extends beneath the linear stability threshold $\Ra_{c,L}$ (Fig.
\ref{fig:bifurc-diag}(a)) to as low as $\Ra=6\times10^5$ $(\approx 0.87 \Ra_c)$ for $\Ha=500$, and
$\Ra=1.7\times10^6$ $(\approx 0.93 \Ra_c)$ for $\Ha=1000$, and is thus subcritical.  These simulations required lengths of approximately 800 or 700 free-fall times (6.5 or 3.4 diffusion times) respectively to assess stability.  Solutions at lower $\Ra$ evolve
extremely slowly, and we have not been able to determine whether they are stable.  
The simplest explanation from the given data is that
this subcritical branch (SB) stems from a saddle-node bifurcation, although we have been unable to determine where the unstable branch from this bifurcation reconnects with the other
branches due to computational cost.  We note that tests conducted using an under-resolved $60^2\times80$ grid did not obtain the correct qualitative transition scenario exhibited by the properly resolved simulations we present here,
and thus, resolving the thin boundary layers on all walls is imperative to assessing the stability of the solutions.
Importantly, the subcritical branch at both magnetic field strengths extends quite close to the conducting state, suggesting that
experimentally small disturbances with the right modal structure could initiate
the transition subcritically.  Although the magnitude of the subcriticality
($\Ra/\Ra_{c,L} \approx 0.9$ in both cases) is quite small, the clear
difference in the symmetries of the solutions is promising and should be
exploited to measure this phenomenon in experiments.  Additionally, the
difference between the lowest $\Ra$ subcritical solution found and $Ra_{c,L}$
grows with increased magnetic field strength by a factor of about 1.5 from
$\Ha=500$ to $\Ha=1000$, suggesting that the gap in $\Ra$ between the point of
linear instability and the point of the expected saddle-node bifurcation grows
with increased magnetic field strength {as approximately $\Ha^{3/2}$. Thus, the subcriticality becomes increasingly important at higher $\Ha$.} 

 \vspace{-10pt}
\subsection{Mixed symmetry branch}
A third branch of equilibrium solutions denoted the mixed symmetry branch (MB) has also been identified, first at $\Ha=1000$ by trying a number of different initial conditions at parameter values where stable equilibria exist.  This branch has then been continued to $\Ha=500$ (Fig. \ref{fig:bifurc-diag}(a)).  The solutions have a symmetry that is mixed between the two other solutions, featuring one/two roll combinations on adjacent walls with matched symmetry on opposing walls (Fig. \ref{fig:bifurc-diag}(b) (MB)).  The vertical velocity field is invariant under the 180$\degree$ rotation $rr_-: (rr_-)w(x,y,z) = -\mathcal{R}_{\pi} w(x,y,-z)$, a reflection in $y$, and a reflection in $x$ accompanied by a reflection in sign and vertical coordinate.  Thus, the symmetry group is isomorphic to the Klein 4-group $K_4 = \{\mathrm{id}, rr_-, sr^3, sr_-\}$.  These stable solutions are interesting as they suggest that narrow/wide rolls characteristic of the linear onset/subcritical branches respectively may coexist.
This is relevant to larger aspect ratio domains where recent simulations from \cite{wu2025flow} show near-onset solutions that have various combinations of wide/narrow rolls.  This suggests that the results are unlikely to be specific to this geometry or aspect ratio.

\vspace{-10pt}

\subsection{Amplitude equations from symmetry}
\begin{figure}
    \captionsetup{width=\columnwidth}
    \centering
    \includegraphics[width=0.6\linewidth]{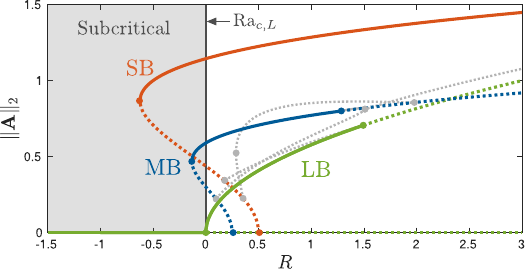}
    \caption{Bifurcation diagram of the amplitude equations \ref{eq:amplitude_sys}, showing the norm of the amplitudes $\|\boldsymbol{A}\|_2$ as a function of the reduced Rayleigh number $R$ for the various stable/unstable equilibria denoted by solid/dotted lines.  Single mode solutions corresponding to the LB, MB and SB states are shown in green, blue and orange respectively.  Mixed mode solutions are shown in grey.  Markers show the bifurcation points.}
    \label{fig:amplitude_eq}
\end{figure}

A model amplitude system which captures the main features of the bifurcation diagram (Fig. \ref{fig:bifurc-diag}(a)), may be deduced from symmetries of the system.  Associated with each of the three equilibrium solutions found, a secondary solution can be obtained by reversing the direction of the rolls and flipping in the vertical direction $z\mapsto -z$.  As $\Ra$ is increased, LB first bifurcates from the conducting state through a \emph{supercritical} pitchfork bifurcation.  We then suppose that MB and SB subsequently bifurcate through \emph{subcritical} pitchfork bifurcations as $\Ra$ is increased further.  Close to onset, when the reduced Rayleigh number $R \sim (\Ra - \Ra_{c,L})/\Ra_{c,L} \ll 1$, we assume that only these three modes are relevant to the dynamics of the system, which can thus be described by three amplitudes $A_n(t)$ corresponding to the LB, SB, and MB ($n={1,2,3}$) equilibria respectively
\vspace{-6pt}
\begin{equation}
    \boldsymbol{\Psi}(x,y,z,t) = \sum_{n=1}^3 \boldsymbol{\Psi}_n(x,y,z,t) + \mathrm{s.m.} = \sum_{n=1}^3 A_n(t) \boldsymbol{\psi}_n(x,y,z) + \mathrm{s.m.}, 
\end{equation}
where $\boldsymbol{\Psi}$ is the state vector of the system and $\mathrm{s.m.}$ are stable modes which will be neglected.  Since for each solution $\boldsymbol{\Psi}_n(x,y,z,t)$, there exists a distinct antisymmetric solution $\sz \boldsymbol{\Psi}_n(x,y,z,t) =  -\boldsymbol{\Psi}_n(x,y,-z,t)$, the equation that governs the evolution of the amplitudes must respect this symmetry.  Thus, writing a general nonlinear evolution equation for the amplitudes as $\partial_t{\boldsymbol{A}} = \mathcal{N}(\boldsymbol{A})$, where $\boldsymbol{A} = [A_1, A_2, A_3]^T$, we note that $\mathcal{N}$ must be equivariant with respect to the reflection $\varpi\boldsymbol{A} \equiv -\boldsymbol{A}$, meaning the commutative property $(\varpi \mathcal{N})(\boldsymbol{A}) = (\mathcal{N}\varpi)(\boldsymbol{A)}$ must hold \citep{crawford1991symmetry}.
This implies that $\mathcal{N}$ is odd $-\mathcal{N}(\boldsymbol{A}) = \mathcal{N}(-\boldsymbol{A})$, and thus, the evolution equation must have the form $\partial_t{\boldsymbol{A}} = \boldsymbol{\mathcal{F}}(A_k^2,R)\boldsymbol{A}$, for $k=\{1,2,3\}$.  We note that these equations are also equivariant with respect to the additional symmetries of the different states.  The information about these symmetries are contained in the eigenfunctions $\boldsymbol{\psi}_n$ and do not change the amplitudes themselves.  We then Taylor expand $\boldsymbol{\mathcal{F}}$, for $R\ll1$, and truncate the system to the lowest order that qualitatively captures the desired dynamics, obtaining 
\begin{subequations}
\label{eq:amplitude_sys}
\begin{alignat}{2}
   \partial_t{A}_1 &= [R - b_1A_2^2 - c_1A_3^2 - d_1A_1^2]A_1,  &(LB)\\
    \partial_t{A}_2 &= [(R-a_2) - b_2A_1^2 - c_2A_3^2 + d_2A_2^2 - e_2A_2^4]A_2, \qquad&(SB) \\
    \partial_t{A}_3 &= [(R-a_3) - b_3A_1^2 - c_3A_2^2 + d_3A_3^2 - e_3A_3^4]A_3, &(MB)
\end{alignat}    
\end{subequations}
where $a_n, b_n, c_n, d_n, e_n>0$ are undetermined real coefficients.  The position of the subcritical pitchfork bifurcations of the SB and MB branches are set by the coefficients $a_2$ and $a_3$.  The position of the saddle-node bifurcations of the SB and MB branches occur at $R_{sn,n} = a_n - d_n^2/4e_n$ at amplitudes $A_n = \pm \sqrt{d_n/2e_n}$ for $n=\{2,3\}$.  The coupling coefficients $b_i, c_i$ determine the bifurcation points, stability and domains of existence of the mixed-mode solutions.  For illustration, we have set $a_2=0.5, a_3=0.25, b_1 = 1.5, e_2=b_2=c_n=2, d_2=3, d_3 = 3.5, b_3=6,e_3=8$.  The bifurcation diagram for this choice or parameters is shown in figure \ref{fig:amplitude_eq}, exhibiting good qualitative agreement with the bifurcation diagram of the full system in figure \ref{fig:bifurc-diag}(a).  Thus, although additional complexity may exist in the full system, the behaviour we observe in the DNS can be rationalised by a simplified system, derived from assumptions about the origin of the SB and MB branches, and a consideration of the symmetries of the problem.

A natural question that arises is whether this behaviour is also expected in a cylindrical geometry, often used in experiments, where the system has a continuous rotational symmetry i.e. $O(2)\times\mathbb{Z}_2$ instead of the discrete $D_4\times\mathbb{Z}_2$ as before.  We describe the competition between $N$ unstable modes of azimuthal wavenumber $n$ in cylindrical coordinates by expanding the state vector as 
\vspace{-6pt}
\begin{equation}
    \boldsymbol{\Psi}(r,\theta,z,t) =  \mathfrak{R}\Big[\sum_{n=1}^N a_n(t) e^{\mathrm{i}n\theta}\boldsymbol{\psi}_n(r,z)\Big] + \mathrm{s.m.}, 
\end{equation}
where $\mathfrak{R[\cdot]}$ takes the real part of the now complex eigenfunction decomposition.  Again, the equation evolving the amplitudes $a_n$ must be equivariant with respect to the symmetries of the system, and thus, must commute with rotations $\theta \rightarrow \theta + \theta_0$ and the reflection.  Thus, $\partial_t{a}_n = \mathcal{F}(|a_k|^2,R)a_n$, for $k\in\{1,\dots,N\}$.  Since the conducting state now bifurcates through a circle pitchfork bifurcation, meaning the solutions are neutrally stable with respect to azimuthal rotations, we write the complex amplitudes in polar form $a_n = A_n e^{\mathrm{i}\varphi_n}$, noting that due to this neutrality, $\partial_t{\varphi}_n=0$.  Again, Taylor expanding $\mathcal{N}$, for $R\ll1$, we arrive at a system for the real amplitudes $A_n$ which is equivalent to Eq. (\ref{eq:amplitude_sys}) for $N$ equations.  This means that the continuous symmetry of a cylinder does not preclude the possibility of having the same behaviour we have observed in the cube.  
Whether the same behaviour occurs as in the cylinder depends on the sign of the coefficients as described before, which must be determined to make qualitative statements about the behaviour of the system.  However, we observe that the continuous symmetry of the cylinder does not affect the final structure of the equations we obtain compared to the cube.  Furthermore, any finite subgroup of $O(2)$ is isomorphic to a dihedral $D_n$ or cyclic $\mathbb{Z}_n$ group and thus, the solutions bifurcating from the conducting state through these circle pitchforks will generically have $D_n$ or $\mathbb{Z}_n$ symmetries.  This suggests that similar symmetry breaking phenomena observed in the cube, may plausibly occur in the cylinder.

\begin{figure}
\captionsetup{width=\columnwidth}
    \includegraphics[width = 1.0\columnwidth]{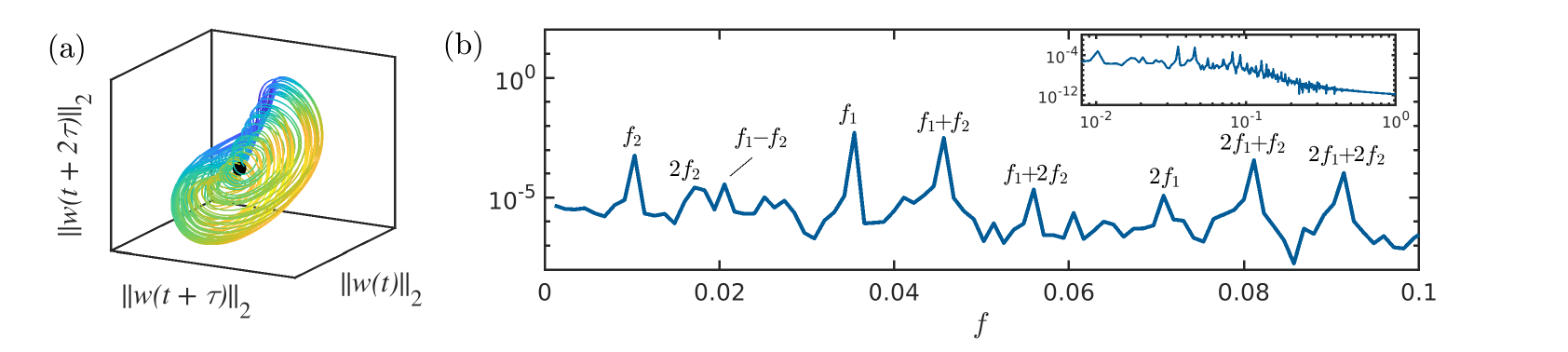}
    \includegraphics[width = 1.0\columnwidth]{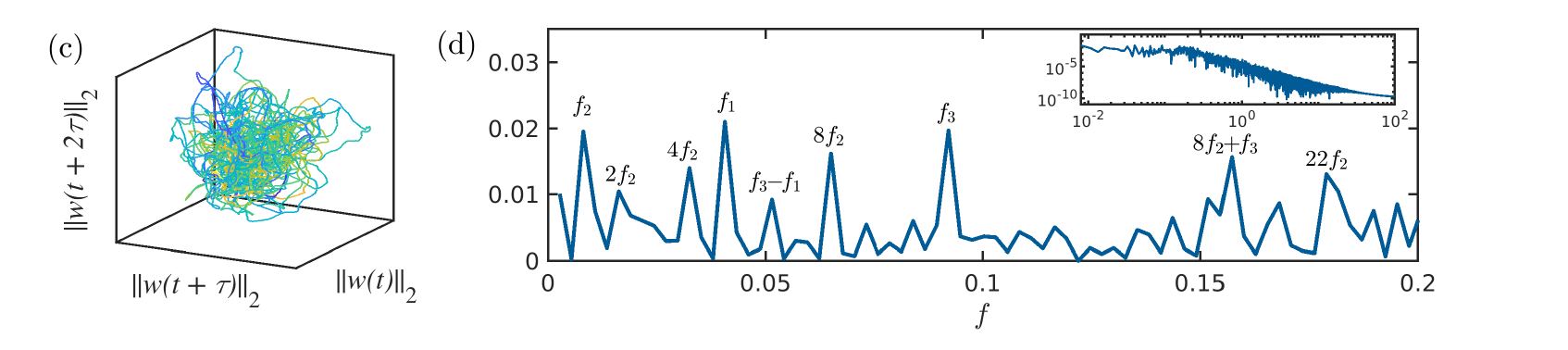}
    \caption{Phase portrait at $\Ha=500$ of the (a) invariant 2-torus at $\Ra=4\times10^6$ and (c) chaotic solution at $\Ra=10^8$ constructed using time-delay embedding with $\tau\approx4$.  The colour map corresponds to $\|w(t+3\tau)\|_2$.  The unstable equilibrium point (LB) is shown in black.  (b,d) Corresponding power spectral density of $\|w(t)\|_2$ respectively as a function of the frequency $f$.  Fundamental frequencies are labelled $f_i$.  Zoomed out spectra are shown in the insets.}
    \label{fig:phase_port_spec}
\end{figure}

 \vspace{-10pt}
\section{Subcritical transition to turbulence}
Although the presence of subcritical solutions is of fundamental interest, the dynamical role that these solutions play in the transition to chaos must be investigated.  Thus, we assess the route to chaos for all three branches at $\Ha=500$.   As previously mentioned, the equilibria on the linear onset branch (LB) become unstable between $3\times 10^6 < \Ra \leq 4\times10^6$, and undergo long transients of approximately 700 free-fall times before ultimately being attracted to solutions on the subcritical branch.  {The mixed symmetry branch (MB) similarly becomes unstable with increased $\Ra$.}  The subcritical branch first undergoes a Hopf bifurcation as $\Ra$ is increased between $3
\times 10^6 < \Ra < 3.1\times10^6$ which breaks the $s$ reflection symmetry, producing a limit cycle featuring a
periodic flapping of four $r_-$ rotationally symmetric wall mode extrusions in the horizontal plane.  This is equivalent to the secondary bifurcation
previously observed at $\Ha=1000$ \citep{McCormack_Teimurazov_Shishkina_Linkmann_2023}.
A secondary Hopf bifurcation (Neimark-Sacker bifurcation on the Poincar\'e section of the limit
cycle) subsequently occurs upon increased $\Ra$, forming an invariant 2-torus.  The invariant 2-torus is characterised by two fundamental frequencies {($f_1 \approx 0.03542$ and $f_2\approx 0.01028$ at $\Ra=4\times10^6$)} which are incommensurable (\emph{i.e.} not rationally related), and thus, the solution is not periodic and densely fills $\mathbb{T}^2$.  This property is confirmed by constructing an arbitrary time series using $f_1$ and $f_2$ in a Fourier expansion, the phase portrait of which is seen to become increasingly dense as time is made arbitrarily large.
The power density spectrum of the solution at $\Ra=4\times10^6$ is shown in Fig. \ref{fig:phase_port_spec}(b) with additional peaks appearing as harmonics of the fundamental frequencies.  The phase portrait of the solution is 
shown in Fig.
\ref{fig:phase_port_spec}(a).  The solution continues to feature flapping wall modes, but the additional frequency modulates the protrusions as they flap, moving their tips back and forth from each other, maintaining the $\mathbb{Z}_4=\{\mathrm{id},r_-\}$ symmetry despite the additional wall mode modulation. 
Increasing the Rayleigh number to
$\Ra=5\times10^6$ leads to a $\mathbb{Z}_4$ to $\mathbb{Z}_2=\{\mathrm{id}, r^2\}$ symmetry break.
Here, the solution still features four
wall modes flapping back and forth as before, but as the wall mode dynamics become more vigorous, extending further into the bulk, the solution now has a
preferred diagonal where two opposing wall modes dominate over the others.
The solution becomes
chaotic by $\Ra=10^7$ and is accompanied by a large increase in fast timescale activity,
revealing a broadband spectrum.  Clear trends are more easily observed at
$\Ra=10^8$, whose power density spectrum is shown in Fig.
\ref{fig:phase_port_spec}(d) and phase portrait is constructed in Fig.
\ref{fig:phase_port_spec}(c).  The solution features three incommensurate
frequencies and a wide broadband spectrum, which can be compared to the
spectrum of the 2-torus at $\Ra=4\times10^6$ in Fig.
\ref{fig:phase_port_spec}(b).  Thus, the system undergoes a quasiperiodic route to chaos where a finite number of Hopf bifurcations lead to the formation of a chaotic attractor.  This is in the spirit of the Ruelle-Takens-Newhouse theorem \citep{ruelle1971nature,newhouse1978occurrence,eckmann1981roads}, which predicts more generally that three-tori are likely to be structurally unstable.  However, more detailed analysis is required to determine precise details of the route in the present system, due to the possibility of phase-locking (tori-periodic orbit transitions) \citep{arnold_1961,Reichhardt_1999}, periodic-chaotic-periodic transitions \citep{Knobloch_Moore_Toomre_Weiss_1986} or other effects.  Exploring the behaviour of the system at finer increments of increased nonlinearity ($\Ra$), as has been done in other fluid systems \citep{Knobloch_Moore_Toomre_Weiss_1986,oteski2015quasiperiodic,PhysRevFluids.4.044401,wang2025feigenbaum}, may well reveal additional complexity.

At $\Ha=1000$, the wall modes display smaller length scales and are increasingly confined near the sidewall, resulting in a different transition pathway.  The subcritical branch undergoes a Hopf bifurcation (similar to the $\Ha=500$ case), but then displays complex homoclinic behaviour reminiscent of Shilnikov dynamics \citep{McCormack_Teimurazov_Shishkina_Linkmann_2023}.  Importantly, the transition again occurs on the subcritical branch and has no connection to the initial linear instability.

\begin{figure}
    \centering
    \captionsetup{width=\columnwidth}
    \includegraphics[width=0.85\linewidth]{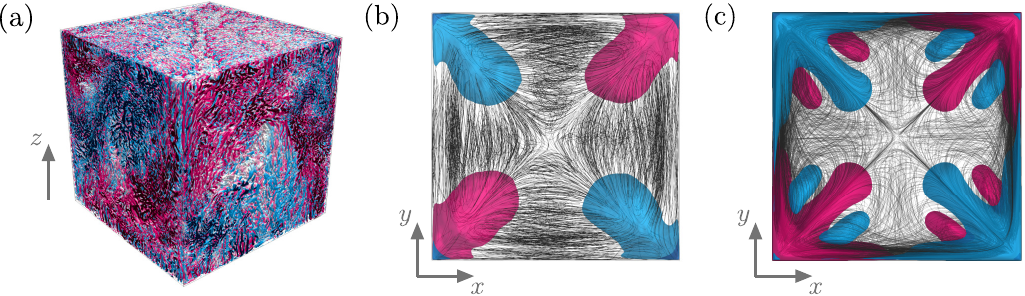}
    \caption{{(a) Instantaneous Q-criterion isosurfaces (Q=3) coloured by the vertical vorticity for the flow at $\Ha=500$, $\Ra=10^9$ and (b) the corresponding mean flow ($w=\pm0.1$ isosurfaces (pink/blue)). (c) Equilibrium solution on the subcritical branch (SB) at $\Ra=2\times 10^6$ ($w=\pm0.01$ isosurfaces).}}
    \label{fig:mean_flow_symm}
\end{figure}

 \vspace{-10pt}
\subsection{Consequences of the subcritical transition}
At all values of $\Ha$ studied, the route to chaos originates through a sequence of
bifurcations on the subcritical branch.  Often in supercritical transitions, the equilibrium solution born from the first linear instability in the system imprints the turbulent flow, organising large-scale features (\emph{e.g.} Taylor vortices in turbulent TCF \citep{taylor1923viii,grossmann2016high} or convection rolls in turbulent RBC \citep{chandrasekhar1961hydrodynamic,Busse_1978,ahlers2009heat}).  Although the subcriticality is reasonably small in this system at the considered $\Ha$, it has a significant effect on the flow in the turbulent regime.  At $\Ra=10^9$ ($\approx 1000\,\Ra_{c,L}$), the flow field exhibits clear multiscale behaviour with small-scale vortices being observed in an instantaneous snapshot of the flow in Fig. \ref{fig:mean_flow_symm}(a).  However, unlike the supercritical cases, this turbulent flow has been formed through bifurcations on the subcritical branch and thus, the turbulent flow retains properties of the subcritical equilibria and not the linear instability.  This is clearly observed in the mean flow of the solution which retains the same $D_4^-$ symmetry as the subcritical equilibria, and not the $D_4^+$ symmetry of the solutions born from the linear instability.  This means that understanding the subcritical nature of the transition is imperative to understanding properties of turbulent MC, such as the globally organising flow structures.

 \vspace{-10pt}
\section{Conclusions}
The addition of insulating sidewalls changes the phenomenology of the transition to turbulence in liquid metal magnetoconvection, with stable solutions existing beneath the linear stability threshold.  {This occurs despite the conducting state undergoing a supercritical bifurcation, and the transition proceeding supercritically in a laterally periodic layer.}  Although the magnitude of the subcriticality is relatively small at the studied magnetic field strengths, our results suggest the transition becomes increasingly subcritical at higher magnetic field strengths, and is thus, centrally important in high $\Ha$ regimes currently inaccessible to simulations and experiments.  
Multiple stable equilibrium solutions exist near onset, which are distinguished by their spatial symmetries and have different transport properties.  The qualitative features of the near onset behaviour can be rationalised by low-dimensional amplitude equations, deduced from symmetries of the system.  We have shown that
the solutions born from the linear instability (LB), and the mixed symmetry branch (MB)
quickly become unstable, and solutions are attracted to the subcritical
branch (SB), which is the primary branch involved in the route to chaos.
Thus, we show that both the transition and the formation of global flow structures in turbulent regimes can only be understood through the subcriticality, and have no connection to the linear instability, 
despite linear stability theory predicting a supercritical transition.
This must be considered when laboratory experiments with sidewalls are compared to laterally unbounded geophysical or astrophysical flows.
{Since wall modes are observed in a variety of container geometries and with finite wall conductivity, both numerically and experimentally, we anticipate the subcriticality is insensitive to these properties.}
Wall modes are also well known to form in rotating convection with sidewalls 
\citep{goldstein1993convection,herrmann_busse_1993,ning1993rotating,kuo1993traveling,goldstein1994convection,liu1999nonlinear,liao2006boundary,kunnen2011role,horn2017prograde,favier2020robust,de2020turbulent,zhang2021boundary,wedi2022experimental,de2023robust,zhang2024wall,Ravichandran_Wettlaufer_2024,vasil2024rapidly},
and are robust to varied sidewall geometry \citep{favier2020robust}. Furthermore, wall localised modes are known to form in a variety of simple pattern forming systems on bounded domains, such as the cubic-quintic Swift--Hohenberg equation, which are expected to have wide applicability \citep{verschueren21}.  Thus, one might anticipate that similar results to those obtained here may extend to
a wider class of systems, such as rotating convection or rotating MC.
Further open questions in this direction relate to obtaining a detailed understanding of these effects for domains with continuous symmetries (\emph{e.g.} for MC in a cylinder), and for larger aspect ratio domains.


\backsection[Acknowledgements]{We thank A.~Morozov and G.~Vasil for helpful discussions.  ML thanks the Isaac Newton Institute for Mathematical Sciences, Cambridge, for
support and hospitality during the programme ``Anti-diffusive dynamics: from
sub-cellular to astrophysical scales" (EPSRC grant EP/R014604/1), where work on this paper was undertaken.} 

\backsection[Funding]{This work was supported by the Deutsche Forschungsgemeinschaft (SPP1881 ``Turbulent
Superstructures" and grants  Sh405/20, Sh405/22, Li3694/1), and used the ARCHER2 UK National Supercomputing Service (https://www.archer2.ac.uk) with resources provided by the UK Turbulence Consortium (EPSRC grants EP/R029326/1, EP/X035484/1).}

\backsection[Declaration of interests]{The authors report no conflict of interest.}

  \vspace{-10pt}
\bibliographystyle{jfm}
\bibliography{jfm}

\begin{thebibliography}{84}
\expandafter\ifx\csname natexlab\endcsname\relax\def\natexlab#1{#1}\fi
\def\au#1{#1} \def\ed#1{#1} \def\yr#1{#1}\def\at#1{#1}\def\jt#1{\textit{#1}} \def\bt#1{#1}\def\bvol#1{\textbf{#1}} \def\vol#1{#1} \def\pg#1{#1} \def\publ#1{#1}\def\arxiv#1{#1}\def\org#1{#1}\def\st#1{\textit{#1}}

\bibitem[Ahlers {\em et~al.\/}(2009)Ahlers, Grossmann \& Lohse]{ahlers2009heat}
{\sc \au{Ahlers, G.}, \au{Grossmann, S.} \& \au{Lohse, D.}} \yr{2009}  \at{Heat transfer and large scale dynamics in turbulent {Rayleigh-B{\'e}nard} convection}.  \jt{Rev. Mod. Phys.}  \bvol{81}~(2),  \pg{503}.

\bibitem[Akhmedagaev {\em et~al.\/}(2020)Akhmedagaev, Zikanov, Krasnov \& Schumacher]{akhmedagaev2020turbulent}
{\sc \au{Akhmedagaev, R.}, \au{Zikanov, O.}, \au{Krasnov, D.} \& \au{Schumacher, J.}} \yr{2020}  \at{Turbulent {Rayleigh--B{\'e}nard} convection in a strong vertical magnetic field}.  \jt{J. Fluid Mech.}  \bvol{895},  \pg{R4}.

\bibitem[Arnol'd(1961)]{arnold_1961}
{\sc \au{Arnol'd, V.~I.}} \yr{1961}  \at{Small denominators. {I}. {M}apping the circle onto itself}.  \jt{Izv. Math.}  \bvol{82}.

\bibitem[Avila {\em et~al.\/}(2023)Avila, Barkley \& Hof]{Avila2023}
{\sc \au{Avila, M.}, \au{Barkley, D.} \& \au{Hof, B.}} \yr{2023}  \at{Transition to turbulence in pipe flow}.  \jt{Annu. Rev. Fluid Mech.}  \bvol{55},  \pg{575--602}.

\bibitem[Bhattacharya {\em et~al.\/}(2024)Bhattacharya, Boeck, Krasnov \& Schumacher]{bhattacharya2024wall}
{\sc \au{Bhattacharya, S.}, \au{Boeck, T.}, \au{Krasnov, D.} \& \au{Schumacher, J.}} \yr{2024}  \at{Wall-attached convection under strong inclined magnetic fields}.  \jt{J. Fluid Mech.}  \bvol{979},  \pg{A53}.

\bibitem[Bodenschatz {\em et~al.\/}(2000)Bodenschatz, Pesch \& Ahlers]{bodenschatz2000rbc}
{\sc \au{Bodenschatz, E.}, \au{Pesch, W.} \& \au{Ahlers, G.}} \yr{2000}  \at{Recent developments in {Rayleigh--B{\'e}nard} convection}.  \jt{Annu. Rev. Fluid Mech.}  \bvol{32},  \pg{709--778}.

\bibitem[Bonn {\em et~al.\/}(2011)Bonn, Ingremeau, Amarouchene \& Kellay]{Bonn2011}
{\sc \au{Bonn, D.}, \au{Ingremeau, F.}, \au{Amarouchene, Y.} \& \au{Kellay, H.}} \yr{2011}  \at{{Large velocity fluctuations in small-Reynolds-number pipe flow of polymer solutions}}.  \jt{Phys. Rev. E}  \bvol{84}~(4, 2).

\bibitem[Brandstater \& Swinney(1987)]{taylorCouette-swinney-1987}
{\sc \au{Brandstater, A.} \& \au{Swinney, H.}} \yr{1987}  \at{Strange attractors in weakly turbulent {Couette--Taylor} flow}.  \jt{Phys. Rev. A}  \bvol{35}~(5),  \pg{2207}.

\bibitem[Busse(1978)]{Busse_1978}
{\sc \au{Busse, F~H}} \yr{1978}  \at{Non-linear properties of thermal convection}.  \jt{Rep. Prog. Phys.}  \bvol{41}~(12),  \pg{1929}.

\bibitem[Busse(2008)]{busse2008asymptotic}
{\sc \au{Busse, F.~H.}} \yr{2008}  \at{Asymptotic theory of wall-attached convection in a horizontal fluid layer with a vertical magnetic field}.  \jt{Phys. Fluids}  \bvol{20}~(2),  \pg{024102}.

\bibitem[Busse \& Clever(1982)]{Busse1982}
{\sc \au{Busse, F.~H.} \& \au{Clever, R.~M.}} \yr{1982}  \at{Stability of convection rolls in the presence of a vertical magnetic field}.  \jt{Phys. Fluids}  \bvol{25}~(6),  \pg{931--935}.

\bibitem[Chandrasekhar(1961)]{chandrasekhar1961hydrodynamic}
{\sc \au{Chandrasekhar, S.}} \yr{1961} {\em Hydrodynamic and Hydromagnetic Stability\/}.  \publ{Oxford University Press}.

\bibitem[Chossat \& Iooss(1994)]{taylorCouetteBook}
{\sc \au{Chossat, P.} \& \au{Iooss, G.}} \yr{1994} {\em The {Couette--Taylor} problem\/},  \st{Applied mathematical sciences},  \vol{vol. 102}.  \publ{Springer-Verlag}.

\bibitem[Crawford \& Knobloch(1991)]{crawford1991symmetry}
{\sc \au{Crawford, J.~D.} \& \au{Knobloch, E.}} \yr{1991}  \at{Symmetry and symmetry-breaking bifurcations in fluid dynamics}.  \jt{Annu. Rev. Fluid Mech.}  \bvol{23}~(1),  \pg{341--387}.

\bibitem[Davidson(1999)]{Davidson1999}
{\sc \au{Davidson, P.~A.}} \yr{1999}  \at{{Magnetohydrodynamics in materials processing}}.  \jt{Annu. Rev. Fluid Mech.}  \bvol{31},  \pg{273--300}.

\bibitem[Ecke {\em et~al.\/}(1991)Ecke, Mainieri \& Sullivan]{PhysRevA.44.8103}
{\sc \au{Ecke, R.~E.}, \au{Mainieri, R.} \& \au{Sullivan, T.~S.}} \yr{1991}  \at{Universality in quasiperiodic {R}ayleigh-{B}\'enard convection}.  \jt{Phys. Rev. A}  \bvol{44},  \pg{8103--8118}.

\bibitem[Eckhardt(2018)]{eckhardt2018transition}
{\sc \au{Eckhardt, B.}} \yr{2018}  \at{Transition to turbulence in shear flows}.  \jt{Phys. A}  \bvol{504},  \pg{121--129}.

\bibitem[Eckhardt {\em et~al.\/}(2007)Eckhardt, Schneider, Hof \& Westerweel]{Eckhardt2007}
{\sc \au{Eckhardt, B.}, \au{Schneider, T.M.}, \au{Hof, B.} \& \au{Westerweel, J.}} \yr{2007}  \at{Turbulence transition in pipe flow}.  \jt{Ann. Rev. Fluid Mech.}  \bvol{39},  \pg{447}.

\bibitem[Eckmann(1981)]{eckmann1981roads}
{\sc \au{Eckmann, J-P}} \yr{1981}  \at{Roads to turbulence in dissipative dynamical systems}.  \jt{Rev. Mod. Phys.}  \bvol{53}~(4),  \pg{643}.

\bibitem[Favier {\em et~al.\/}(2019)Favier, Guervilly \& Knobloch]{favier2019subcritical}
{\sc \au{Favier, Benjamin}, \au{Guervilly, C{\'e}line} \& \au{Knobloch, Edgar}} \yr{2019}  \at{Subcritical turbulent condensate in rapidly rotating rayleigh--b{\'e}nard convection}.  \jt{J. Fluid Mech.}  \bvol{864},  \pg{R1}.

\bibitem[Favier \& Knobloch(2020)]{favier2020robust}
{\sc \au{Favier, B.} \& \au{Knobloch, E.}} \yr{2020}  \at{Robust wall states in rapidly rotating {Rayleigh--B{\'e}nard} convection}.  \jt{J. Fluid Mech.}  \bvol{895},  \pg{R1}.

\bibitem[Gallet(2024)]{Gallet2024}
{\sc \au{Gallet, B.}} \yr{2024}  \at{Two-dimensional turbulence above topography: condensation transition and selection of minimum enstrophy solutions}.  \jt{J. Fluid Mech.}  \bvol{988},  \pg{A13}.

\bibitem[Goldstein {\em et~al.\/}(1993)Goldstein, Knobloch, Mercader \& Net]{goldstein1993convection}
{\sc \au{Goldstein, H.~F.}, \au{Knobloch, E}, \au{Mercader, I} \& \au{Net, M}} \yr{1993}  \at{Convection in a rotating cylinder. {P}art 1 linear theory for moderate {P}randtl numbers}.  \jt{J. Fluid Mech.}  \bvol{248},  \pg{583--604}.

\bibitem[Goldstein {\em et~al.\/}(1994)Goldstein, Knobloch, Mercader \& Net]{goldstein1994convection}
{\sc \au{Goldstein, H.~F.}, \au{Knobloch, E}, \au{Mercader, I} \& \au{Net, M}} \yr{1994}  \at{Convection in a rotating cylinder. {P}art 2. linear theory for low {P}randtl numbers}.  \jt{J. Fluid Mech.}  \bvol{262},  \pg{293--324}.

\bibitem[Gollub \& Swinney(1975)]{taylorCouette-swinney-1975}
{\sc \au{Gollub, J.~P.} \& \au{Swinney, H.~L.}} \yr{1975}  \at{Onset of turbulence in a rotating fluid}.  \jt{Phys. Rev. Lett.}  \bvol{35}~(14),  \pg{927}.

\bibitem[Grossmann {\em et~al.\/}(2016)Grossmann, Lohse \& Sun]{grossmann2016high}
{\sc \au{Grossmann, S.}, \au{Lohse, D.} \& \au{Sun, C.}} \yr{2016}  \at{{High--Reynolds} number {Taylor--Couette} turbulence}.  \jt{Annu. Rev. Fluid Mech.}  \bvol{48}~(1),  \pg{53--80}.

\bibitem[Herrmann \& Busse(1993)]{herrmann_busse_1993}
{\sc \au{Herrmann, J.} \& \au{Busse, F.~H.}} \yr{1993}  \at{Asymptotic theory of wall-attached convection in a rotating fluid layer}.  \jt{J.~Fluid Mech.}  \bvol{255},  \pg{183–194}.

\bibitem[Hof(2023)]{Hof2023}
{\sc \au{Hof, B.}} \yr{2023}  \at{Directed percolation and the transition to turbulence}.  \jt{Nat. Rev. Phys.}  \bvol{5},  \pg{62--72}.

\bibitem[Horn \& Schmid(2017)]{horn2017prograde}
{\sc \au{Horn, S.} \& \au{Schmid, P.~J.}} \yr{2017}  \at{Prograde, retrograde, and oscillatory modes in rotating {Rayleigh--B{\'e}nard} convection}.  \jt{J. Fluid Mech.}  \bvol{831},  \pg{182--211}.

\bibitem[Houchens {\em et~al.\/}(2002)Houchens, Witkowski \& Walker]{houchens2002rayleigh}
{\sc \au{Houchens, B.~C.}, \au{Witkowski, L.~M.} \& \au{Walker, J.~S.}} \yr{2002}  \at{{Rayleigh--B{\'e}nard} instability in a vertical cylinder with a vertical magnetic field}.  \jt{J. Fluid Mech.}  \bvol{469},  \pg{189--207}.

\bibitem[Jones(2011)]{Jones2011}
{\sc \au{Jones, C.~A.}} \yr{2011}  \at{{Planetary magnetic fields and fluid dynamos}}.  \jt{Annu. Rev. Fluid Mech.}  \bvol{43},  \pg{583--614}.

\bibitem[Kerswell(2005)]{Kerswell2005}
{\sc \au{Kerswell, R.~R.}} \yr{2005}  \at{Recent progress in understanding the transition to turbulence}.  \jt{Nonlinearity}  \bvol{18},  \pg{R17--R44}.

\bibitem[Knobloch {\em et~al.\/}(1986)Knobloch, Moore, Toomre \& Weiss]{Knobloch_Moore_Toomre_Weiss_1986}
{\sc \au{Knobloch, E.}, \au{Moore, D.~R.}, \au{Toomre, J.} \& \au{Weiss, N.~O.}} \yr{1986}  \at{Transitions to chaos in two-dimensional double-diffusive convection}.  \jt{J. Fluid Mech.}  \bvol{166},  \pg{409–448}.

\bibitem[Kooij {\em et~al.\/}(2018)Kooij, Botchev, Frederix, Geurts, Horn, Lohse, van~der Poel, Shishkina, Stevens \& Verzicco]{kooij2018comparison}
{\sc \au{Kooij, G.~L.}, \au{Botchev, M.~A.}, \au{Frederix, E. M.~A.}, \au{Geurts, B.~J.}, \au{Horn, S.}, \au{Lohse, D.}, \au{van~der Poel, E.~P.}, \au{Shishkina, O.}, \au{Stevens, R. J. A.~M.} \& \au{Verzicco, R.}} \yr{2018}  \at{Comparison of computational codes for direct numerical simulations of turbulent {Rayleigh--B{\'e}nard} convection}.  \jt{Comput. Fluids}  \bvol{166},  \pg{1--8}.

\bibitem[Kunnen {\em et~al.\/}(2011)Kunnen, Stevens, Overkamp, Sun, van Heijst \& Clercx]{kunnen2011role}
{\sc \au{Kunnen, R.}, \au{Stevens, R.}, \au{Overkamp, J.}, \au{Sun, C.}, \au{van Heijst, G.} \& \au{Clercx, H.}} \yr{2011}  \at{The role of {S}tewartson and {E}kman layers in turbulent rotating {R}ayleigh--{B}{\'e}nard convection}.  \jt{J. Fluid Mech.}  \bvol{688},  \pg{422--442}.

\bibitem[Kuo \& Cross(1993)]{kuo1993traveling}
{\sc \au{Kuo, E.Y.} \& \au{Cross, M.C.}} \yr{1993}  \at{Traveling-wave wall states in rotating {Rayleigh--B{\'e}nard} convection}.  \jt{Phys. Rev. E}  \bvol{47}~(4),  \pg{R2245}.

\bibitem[Launay {\em et~al.\/}(2019)Launay, Cambonie, Henry, Poth\'erat \& Botton]{PhysRevFluids.4.044401}
{\sc \au{Launay, G.}, \au{Cambonie, T.}, \au{Henry, D.}, \au{Poth\'erat, A.} \& \au{Botton, V.}} \yr{2019}  \at{Transition to chaos in an acoustically driven cavity flow}.  \jt{Phys. Rev. Fluids}  \bvol{4},  \pg{044401}.

\bibitem[Lellep {\em et~al.\/}(2024)Lellep, Linkmann \& Morozov]{lellep2024elastic}
{\sc \au{Lellep, M.}, \au{Linkmann, M.} \& \au{Morozov, A.}} \yr{2024}  \at{Purely elastic turbulence in pressure-driven channel flows}.  \jt{Proc. Natl. Acad. Sci.}  \bvol{121},  \pg{e2318851121}.

\bibitem[Liao {\em et~al.\/}(2006)Liao, Zhang \& Chang]{liao2006boundary}
{\sc \au{Liao, X}, \au{Zhang, K} \& \au{Chang, Y}} \yr{2006}  \at{On boundary-layer convection in a rotating fluid layer}.  \jt{J. Fluid Mech.}  \bvol{549},  \pg{375--384}.

\bibitem[Libchaber {\em et~al.\/}(1982)Libchaber, Laroche \& Fauve]{libchaber1982period}
{\sc \au{Libchaber, A.}, \au{Laroche, C.} \& \au{Fauve, S.}} \yr{1982}  \at{Period doubling cascade in mercury, a quantitative measurement}.  \jt{J. Physique Lett.}  \bvol{43}~(7),  \pg{211--216}.

\bibitem[Linkmann {\em et~al.\/}(2019)Linkmann, Boffetta, Marchetti \& Eckhardt]{Linkmann2019}
{\sc \au{Linkmann, M.}, \au{Boffetta, G.}, \au{Marchetti, M.~C.} \& \au{Eckhardt, B.}} \yr{2019}  \at{Phase transition to large scale coherent structures in two-dimensional active matter turbulence}.  \jt{Phys. Rev. Lett.}  \bvol{122},  \pg{214503}.

\bibitem[Linkmann {\em et~al.\/}(2020{\natexlab{{\em a\/}}})Linkmann, Hohmann \& Eckhardt]{Linkmann2020b}
{\sc \au{Linkmann, M.}, \au{Hohmann, M.} \& \au{Eckhardt, B.}} \yr{2020{\natexlab{{\em a\/}}}}  \at{Non-universal transitions to two-dimensional turbulence}.  \jt{J. Fluid Mech.}  \bvol{892},  \pg{A18}.

\bibitem[Linkmann {\em et~al.\/}(2020{\natexlab{{\em b\/}}})Linkmann, Marchetti, Boffetta \& Eckhardt]{Linkmann2020a}
{\sc \au{Linkmann, M.}, \au{Marchetti, M.~C.}, \au{Boffetta, G.} \& \au{Eckhardt, B.}} \yr{2020{\natexlab{{\em b\/}}}}  \at{Condensate formation and multiscale dynamics in two-dimensional active suspensions}.  \jt{Phys. Rev. E}  \bvol{101},  \pg{022609}.

\bibitem[Linkmann \& Morozov(2015)]{Linkmann2015}
{\sc \au{Linkmann, M.~F.} \& \au{Morozov, A.}} \yr{2015}  \at{Sudden relaminarization and lifetimes in forced isotropic turbulence}.  \jt{Phys. Rev. Lett.}  \bvol{115},  \pg{134502}.

\bibitem[Liu {\em et~al.\/}(2018)Liu, Krasnov \& Schumacher]{Liu2018}
{\sc \au{Liu, W.}, \au{Krasnov, D.} \& \au{Schumacher, J.}} \yr{2018}  \at{Wall modes in magnetoconvection at high {H}artmann numbers}.  \jt{J. Fluid Mech.}  \bvol{849},  \pg{R21--R212}.

\bibitem[Liu \& Ecke(1999)]{liu1999nonlinear}
{\sc \au{Liu, Y.} \& \au{Ecke, R.~E.}} \yr{1999}  \at{Nonlinear traveling waves in rotating {Rayleigh--B{\'e}nard} convection: Stability boundaries and phase diffusion}.  \jt{Phys. Rev. E}  \bvol{59}~(4),  \pg{4091}.

\bibitem[Lohse \& Shishkina(2023)]{lohse2023ultimate}
{\sc \au{Lohse, D.} \& \au{Shishkina, O.}} \yr{2023}  \at{{Ultimate turbulent thermal convection}}.  \jt{Phys. Today}  \bvol{76}~(11),  \pg{26--32}.

\bibitem[Lohse \& Shishkina(2024)]{lohse2024ultimate}
{\sc \au{Lohse, D.} \& \au{Shishkina, O.}} \yr{2024}  \at{Ultimate {Rayleigh--B{\'e}nard} turbulence}.  \jt{Rev. Mod. Phys.}  \bvol{96}~(3),  \pg{035001}.

\bibitem[Malkus \& Veronis(1958)]{Malkus_Veronis_1958}
{\sc \au{Malkus, W. V.~R.} \& \au{Veronis, G.}} \yr{1958}  \at{{F}inite amplitude cellular convection}.  \jt{J. Fluid Mech.}  \bvol{4}~(3),  \pg{225–260}.

\bibitem[McCormack {\em et~al.\/}(2023)McCormack, Teimurazov, Shishkina \& Linkmann]{McCormack_Teimurazov_Shishkina_Linkmann_2023}
{\sc \au{McCormack, M.}, \au{Teimurazov, A.}, \au{Shishkina, O.} \& \au{Linkmann, M.}} \yr{2023}  \at{Wall mode dynamics and transition to chaos in magnetoconvection with a vertical magnetic field}.  \jt{J. Fluid Mech.}  \bvol{975},  \pg{R2}.

\bibitem[Morozov \& van Saarloos(2007)]{Morozov2007}
{\sc \au{Morozov, A.~N.} \& \au{van Saarloos, W.}} \yr{2007}  \at{An introductory essay on subcritical instabilities and the transition to turbulence in visco-elastic parallel shear flows}.  \jt{Phys. Rep.}  \bvol{447},  \pg{112--143}.

\bibitem[Newhouse {\em et~al.\/}(1978)Newhouse, Ruelle \& Takens]{newhouse1978occurrence}
{\sc \au{Newhouse, S.}, \au{Ruelle, D.} \& \au{Takens, F.}} \yr{1978}  \at{Occurrence of strange axiom a attractors near quasi periodic flows on t m, m$\geq$ 3}.  \jt{Comm. Math. Phys.}  \bvol{64},  \pg{35--40}.

\bibitem[Boro\ifmmode~\acute{n}\else \'{n}\fi{}ska \& Tuckerman(2010)]{PhysRevE.81.036321}
{\sc \au{Boro\ifmmode~\acute{n}\else \'{n}\fi{}ska, K.} \& \au{Tuckerman, L.~S.}} \yr{2010}  \at{Extreme multiplicity in cylindrical {R}ayleigh-{B}\'enard convection. {II}. {B}ifurcation diagram and symmetry classification}.  \jt{Phys. Rev. E}  \bvol{81},  \pg{036321}.

\bibitem[Ning \& Ecke(1993)]{ning1993rotating}
{\sc \au{Ning, L.} \& \au{Ecke, R.~E.}} \yr{1993}  \at{Rotating {R}ayleigh-{B}{\'e}nard convection: Aspect-ratio dependence of the initial bifurcations}.  \jt{Phys. Rev. E}  \bvol{47}~(5),  \pg{3326}.

\bibitem[Oteski {\em et~al.\/}(2015)Oteski, Duguet, Pastur \& Le~Qu{\'e}r{\'e}]{oteski2015quasiperiodic}
{\sc \au{Oteski, L.}, \au{Duguet, Y.}, \au{Pastur, L.} \& \au{Le~Qu{\'e}r{\'e}, P.}} \yr{2015}  \at{Quasiperiodic routes to chaos in confined two-dimensional differential convection}.  \jt{Phys. Rev. E}  \bvol{92}~(4),  \pg{043020}.

\bibitem[Pan {\em et~al.\/}(2013)Pan, Morozov, Wagner \& Arratia]{Pan2013}
{\sc \au{Pan, L.}, \au{Morozov, A.}, \au{Wagner, C.} \& \au{Arratia, P.~E.}} \yr{2013}  \at{Nonlinear elastic instability in channel flows at low {R}eynolds numbers}.  \jt{Phys. Rev. Lett.}  \bvol{110},  \pg{174502}.

\bibitem[Proctor \& Weiss(1982)]{proctor1982magnetoconvection}
{\sc \au{Proctor, MRE} \& \au{Weiss, NO}} \yr{1982}  \at{Magnetoconvection}.  \jt{Rep. Prog. Phys.}  \bvol{45}~(11),  \pg{1317}.

\bibitem[Ravichandran \& Wettlaufer(2024)]{Ravichandran_Wettlaufer_2024}
{\sc \au{Ravichandran, S.} \& \au{Wettlaufer, J.S.}} \yr{2024}  \at{Prograde and meandering wall modes in rotating {Rayleigh--B{\'e}nard} convection with conducting walls}.  \jt{J. Fluid Mech.}  \bvol{998},  \pg{A47}.

\bibitem[Rayleigh(1916)]{rayleigh1916lix}
{\sc \au{Rayleigh, Lord}} \yr{1916}  \at{{LIX}. {O}n convection currents in a horizontal layer of fluid, when the higher temperature is on the under side}.  \jt{Lond. Edinb. Dublin Philos. Mag. J. Sci}  \bvol{32}~(192),  \pg{529--546}.

\bibitem[Reichhardt \& Nori(1999)]{Reichhardt_1999}
{\sc \au{Reichhardt, C.} \& \au{Nori, F.}} \yr{1999}  \at{Phase locking, devil's staircases, farey trees, and arnold tongues in driven vortex lattices with periodic pinning}.  \jt{Phys. Rev. Lett.}  \bvol{82},  \pg{414--417}.

\bibitem[Reiter {\em et~al.\/}(2021)Reiter, Shishkina, Lohse \& Krug]{reiter2021crossover}
{\sc \au{Reiter, P.}, \au{Shishkina, O.}, \au{Lohse, D.} \& \au{Krug, D.}} \yr{2021}  \at{Crossover of the relative heat transport contributions of plume ejecting and impacting zones in turbulent {Rayleigh-B{\'e}nard} convection (a)}.  \jt{Europhys. lett.}  \bvol{134}~(3),  \pg{34002}.

\bibitem[Reiter {\em et~al.\/}(2022)Reiter, Zhang \& Shishkina]{reiter2022flow}
{\sc \au{Reiter, P.}, \au{Zhang, X.} \& \au{Shishkina, O.}} \yr{2022}  \at{Flow states and heat transport in {Rayleigh--B{\'e}nard} convection with different sidewall boundary conditions}.  \jt{J. Fluid Mech.}  \bvol{936},  \pg{A32}.

\bibitem[Roche(2020)]{Roche2020}
{\sc \au{Roche, P.~E.}} \yr{2020}  \at{The ultimate state of convection: a unifying picture of very high {R}ayleigh numbers experiments}.  \jt{New J. Phys.}  \bvol{22}~(7),  \pg{073056}.

\bibitem[Ruelle \& Takens(1971)]{ruelle1971nature}
{\sc \au{Ruelle, D.} \& \au{Takens, F.}} \yr{1971}  \at{On the nature of turbulence}.  \jt{Comm. Math. Phys.}  \bvol{20},  \pg{167--192}.

\bibitem[Shishkina \& Lohse(2024)]{shishkina2024ultimate}
{\sc \au{Shishkina, O.} \& \au{Lohse, D.}} \yr{2024}  \at{Ultimate regime of {R}ayleigh-{B}\'enard turbulence: Subregimes and their scaling relations for the {N}usselt vs {R}ayleigh and {P}randtl numbers}.  \jt{Phys. Rev. Lett.}  \bvol{133},  \pg{144001}.

\bibitem[Stefani(2024)]{stefani2024liquid}
{\sc \au{Stefani, F.}} \yr{2024}  \at{Liquid-metal experiments on geophysical and astrophysical phenomena}.  \jt{Nat. Rev. Phys.}  \pg{pp. 1--17}.

\bibitem[Taylor(1923)]{taylor1923viii}
{\sc \au{Taylor, G.~I.}} \yr{1923}  \at{{VIII}. {S}tability of a viscous liquid contained between two rotating cylinders}.  \jt{Philos. Trans. R. Soc. A}  \bvol{223}~(605-615),  \pg{289--343}.

\bibitem[Teimurazov {\em et~al.\/}(2024)Teimurazov, McCormack, Linkmann \& Shishkina]{Teimurazov_2024}
{\sc \au{Teimurazov, A.}, \au{McCormack, M.}, \au{Linkmann, M.} \& \au{Shishkina, O.}} \yr{2024}  \at{Unifying heat transport model for the transition between buoyancy-dominated and {L}orentz-force-dominated regimes in quasistatic magnetoconvection}.  \jt{J. Fluid Mech.}  \bvol{980},  \pg{R3}.

\bibitem[Vasil {\em et~al.\/}(2025)Vasil, Burns, Lecoanet, Oishi, Brown \& Julien]{vasil2024rapidly}
{\sc \au{Vasil, G.~M.}, \au{Burns, K.~J.}, \au{Lecoanet, D.}, \au{Oishi, J.~S.}, \au{Brown, B.} \& \au{Julien, K.}} \yr{2025}  \at{Rapidly rotating wall-mode convection}.  \jt{J. Fluid Mech.}  \bvol{1017},  \pg{A37}.

\bibitem[Verschueren {\em et~al.\/}(2021)Verschueren, Knobloch \& Uecker]{verschueren21}
{\sc \au{Verschueren, N.}, \au{Knobloch, E.} \& \au{Uecker, H.}} \yr{2021}  \at{Localized and extended patterns in the cubic-quintic swift-hohenberg equation on a disk}.  \jt{Phys. Rev. E}  \bvol{104},  \pg{014208}.

\bibitem[Wang {\em et~al.\/}(2025)Wang, Ayats, Deguchi, Meseguer \& Mellibovsky]{wang2025feigenbaum}
{\sc \au{Wang, Baoying}, \au{Ayats, Roger}, \au{Deguchi, Kengo}, \au{Meseguer, Alvaro} \& \au{Mellibovsky, Fernando}} \yr{2025}  \at{Feigenbaum universality in subcritical {Taylor--Couette} flow}.  \jt{J. Fluid Mech.}  \bvol{1010},  \pg{A36}.

\bibitem[Wedi {\em et~al.\/}(2022)Wedi, Moturi, Funfschilling \& Weiss]{wedi2022experimental}
{\sc \au{Wedi, M.}, \au{Moturi, V.~M.}, \au{Funfschilling, D.} \& \au{Weiss, S.}} \yr{2022}  \at{Experimental evidence for the boundary zonal flow in rotating {Rayleigh--B{\'e}nard} convection}.  \jt{J. Fluid Mech.}  \bvol{939},  \pg{A14}.

\bibitem[Weiss \& Proctor(2014)]{weiss2014magnetoconvection}
{\sc \au{Weiss, N.~O.} \& \au{Proctor, M. R.~E.}} \yr{2014} {\em Magnetoconvection\/}.  \publ{Cambridge University Press}.

\bibitem[de~Wit {\em et~al.\/}(2020)de~Wit, Aguirre~Guzm{\'a}n, Madonia, Cheng, Clercx \& Kunnen]{de2020turbulent}
{\sc \au{de~Wit, X.~M.}, \au{Aguirre~Guzm{\'a}n, A.~J.}, \au{Madonia, M.}, \au{Cheng, J.~S.}, \au{Clercx, H. J.~H.} \& \au{Kunnen, R. P.~J.}} \yr{2020}  \at{Turbulent rotating convection confined in a slender cylinder: the sidewall circulation}.  \jt{Phys. Rev. Fluids}  \bvol{5}~(2),  \pg{023502}.

\bibitem[de~Wit {\em et~al.\/}(2022{\natexlab{{\em a\/}}})de~Wit, Aguirre~Guzmán, Clercx \& Kunnen]{deWit2022b}
{\sc \au{de~Wit, Xander~M.}, \au{Aguirre~Guzmán, Andrés~J.}, \au{Clercx, Herman~J.H.} \& \au{Kunnen, Rudie~P.J.}} \yr{2022{\natexlab{{\em a\/}}}}  \at{Discontinuous transitions towards vortex condensates in buoyancy-driven rotating turbulence}.  \jt{J. Fluid Mech.}  \bvol{936},  \pg{A43}.

\bibitem[de~Wit {\em et~al.\/}(2023)de~Wit, Boot, Madonia, Aguirre~Guzm{\'a}n \& Kunnen]{de2023robust}
{\sc \au{de~Wit, X.~M.}, \au{Boot, W. J.~M.}, \au{Madonia, M.}, \au{Aguirre~Guzm{\'a}n, A.~J.} \& \au{Kunnen, R. P.~J.}} \yr{2023}  \at{Robust wall modes and their interplay with bulk turbulence in confined rotating {R}ayleigh-{B}{\'e}nard convection}.  \jt{Phys. Rev. Fluids}  \bvol{8}~(7),  \pg{073501}.

\bibitem[de~Wit {\em et~al.\/}(2022{\natexlab{{\em b\/}}})de~Wit, van Kan \& Alexakis]{deWit2022a}
{\sc \au{de~Wit, Xander~M.}, \au{van Kan, Adrian} \& \au{Alexakis, Alexandros}} \yr{2022{\natexlab{{\em b\/}}}}  \at{Bistability of the large-scale dynamics in quasi-two-dimensional turbulence}.  \jt{J. Fluid Mech.}  \bvol{939},  \pg{R2}.

\bibitem[Wu {\em et~al.\/}(2025)Wu, Chen \& Ni]{wu2025flow}
{\sc \au{Wu, K.}, \au{Chen, L.} \& \au{Ni, M-J.}} \yr{2025}  \at{Flow and heat transfer mechanism of wall mode in {Rayleigh--B{\'e}nard} convection under strong magnetic fields}.  \jt{Phys. Rev. Fluids}  \bvol{10}~(3),  \pg{033702}.

\bibitem[Xu {\em et~al.\/}(2023)Xu, Horn \& Aurnou]{xu2023transition}
{\sc \au{Xu, Y.}, \au{Horn, S.} \& \au{Aurnou, J.~M.}} \yr{2023}  \at{Transition from wall modes to multimodality in liquid gallium magnetoconvection}.  \jt{Phys. Rev. Fluids}  \bvol{8},  \pg{103503}.

\bibitem[Yokoyama \& Takaoka(2017)]{yokoyama2017}
{\sc \au{Yokoyama, N.} \& \au{Takaoka, M.}} \yr{2017}  \at{Hysteretic transitions between quasi-two-dimensional flow and three-dimensional flow in forced rotating turbulence}.  \jt{Phys. Rev. Fluids}  \bvol{2},  \pg{092602}.

\bibitem[Zhang {\em et~al.\/}(2021)Zhang, Ecke \& Shishkina]{zhang2021boundary}
{\sc \au{Zhang, X.}, \au{Ecke, R.~E.} \& \au{Shishkina, O.}} \yr{2021}  \at{Boundary zonal flows in rapidly rotating turbulent thermal convection}.  \jt{J. Fluid Mech.}  \bvol{915},  \pg{A62}.

\bibitem[Zhang {\em et~al.\/}(2024)Zhang, Reiter, Shishkina \& Ecke]{zhang2024wall}
{\sc \au{Zhang, X.}, \au{Reiter, P.}, \au{Shishkina, O.} \& \au{Ecke, R.~E.}} \yr{2024}  \at{Wall modes and the transition to bulk convection in rotating {R}ayleigh-{B}{\'e}nard convection}.  \jt{Phys. Rev. Fluids}  \bvol{9}~(5),  \pg{053501}.

\bibitem[Zikanov {\em et~al.\/}(2014)Zikanov, Krasnov, Boeck, Thess \& Rossi]{Zikanov2014}
{\sc \au{Zikanov, Oleg}, \au{Krasnov, Dmitry}, \au{Boeck, Thomas}, \au{Thess, Andre} \& \au{Rossi, Maurice}} \yr{2014}  \at{Laminar-turbulent transition in magnetohydrodynamic duct, pipe, and channel flows}.  \jt{Appl. Mech. Rev.}  \bvol{66}.

\bibitem[Z{\"u}rner {\em et~al.\/}(2020)Z{\"u}rner, Schindler, Vogt, Eckert \& Schumacher]{zurner2020flow}
{\sc \au{Z{\"u}rner, T.}, \au{Schindler, F.}, \au{Vogt, T.}, \au{Eckert, S.} \& \au{Schumacher, J.}} \yr{2020}  \at{Flow regimes of {Rayleigh--B{\'e}nard} convection in a vertical magnetic field}.  \jt{J. Fluid Mech.}  \bvol{894},  \pg{A21}.

\end{thebibliography}

\end{document}